\documentclass[10pt,letterpaper,preprint]{article} 

\usepackage{ol2}
\usepackage[draft]{hyperref}
\usepackage{amsmath}
\usepackage{graphicx}

\begin{document}


\title{Photonic molecules formed by coupled hybrid resonators}

\author{Bo Peng,$^{1}$ \c{S}ahin Kaya \"{O}zdemir,$^{1,\ast}$ Jiangang Zhu$^{1}$, and Lan Yang $^{1,\ast}$}

\address{
$^1$Department of Electrical and Systems Engineering, Washington University in St. Louis, St. Louis, MO 63130, USA
$^\ast$Corresponding author: yang@ese.wustl.edu, ozdemir@ese.wustl.edu
}

\begin{abstract}We describe a method that enables free-standing whispering-gallery-mode microresonators, and report spectral tuning of photonic molecules formed by coupled free and on-chip resonators with different geometries and materials. We study direct coupling via evanescent fields of free silica microtoroids and microspheres with on-chip polymer coated silica microtoroids. We demonstrate thermal tuning of resonance modes to achieve maximal spectral overlap, mode splitting induced by direct coupling, and the effects of distance between the resonators on the splitting spectra. \end{abstract}


\noindent Whispering-gallery-mode (WGM) optical resonators with their micro-scale mode volumes and high quality factors ($Q$) have been widely used in different areas ranging from sensing and lasing to opto-mechanics \cite{va2004,zhu2010,he2011,kipp2007}. Modification of the structures of the modes in these resonators has been of considerable interest for their potential applications and underlying physics \cite{naw2005,ma2004,little2004,sav2005}. Direct and indirect coupling of two or more resonators have been shown to have impact on their optical properties and to allow manipulation of light field \cite{li2010,xiao2008}. Mode-splitting and coupling efficiency have been demonstrated in coupled microspheres and photonic-crystal cavities \cite{kan2006,ash2006,atl2008}. Anti-crossing of the resonances and $Q$ enhancement of supermodes have been shown in directly coupled GaAs microdisks \cite{ben2011} and silica microtoroids \cite{gru2010}.

Two or more coupled resonators form a compound structure - photonic molecule (PM) - in which interactions of optical modes create supermodes \cite{for2000,rak2010}. This molecular analogy stems from the observation that confined optical modes of a resonator and the electron states of atoms behave similarly. Thus, a single resonator is considered as a "photonic atom", and a pair of coupled resonators as the photonic analog of a molecule. Studying the interactions in PMs is critical to understand their resonance properties and the field and energy transfers to engineer new devices such as phonon lasers \cite {gru2010} and  enhanced sensors \cite{bor2010}.

In this Letter, we introduce a method to detach an on-chip microresonator from its pillar, connecting it to the substrate, to form a free \begin{figure}
\center \includegraphics[width=12.0cm]{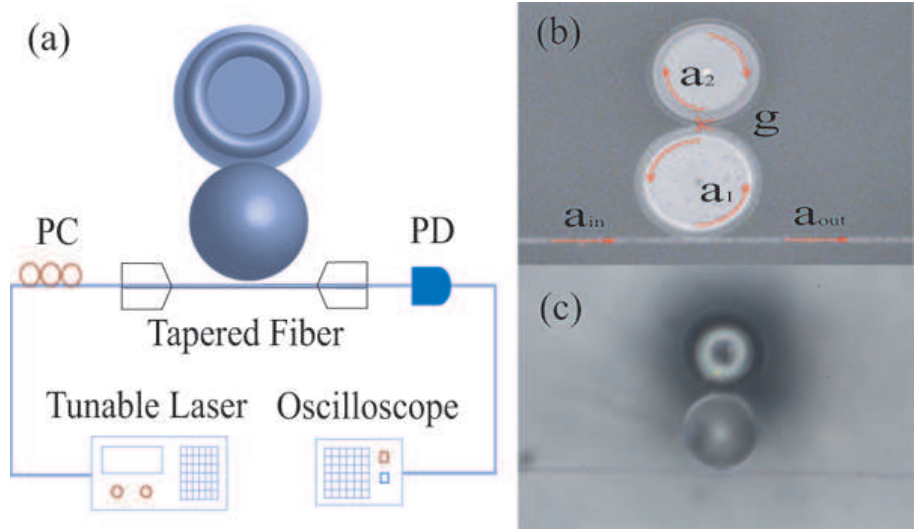}  \caption{ Experimental setup (a) and the optical microscope images of the photonic molecules formed by (b) an on-chip silica microtoroid and a free silica microtoroid, and (c) an on-chip PDMS coated silica microtoroid and a free silica microsphere. PC and PD denote the polarization controller and the photodetector, respectively.\vspace{-1 mm}} \label{Fig1}
\end{figure}
high-$Q$ microresonator, and report studies of the optical modes in PMs made from pairs of directly coupled free and on-chip WGM resonators of different shapes, sizes and materials (hybrid resonators). We investigate the interaction and supermode formation in two types of PMs: (i) Directly coupled on-chip and free silica microtoroids (PM1), and (ii) directly coupled on-chip PDMS coated silica microtoroid and a silica microsphere having a fiber stem (PM2).

The experimental setup and microscope images of the PMs are given in Fig. \ref{Fig1}. Silica sphere and toroids are fabricated according to Refs\cite{cai2001,arm2003}. Some of the silica toroids were coated with a thin layer of PDMS \cite{he2009}. We chose the resonators of a PM with approximately equal radii, $\sim35\mu m$ in PM1 and $\sim30\mu m$ in PM2.

The free silica microtoroid of PM1 is fabricated as follows. First, the radius of the silicon pillar of an on-chip microtoroid is reduced to $\sim 1-2\mu m$. Then a fiber whose tip is cleaved and coated with optical glue is brought to the inner disk of the microtoroid such that the fiber does not touch the sidewalls of the toroid. This ensures that the smooth circular rim of the toroid where the optical mode resides stay intact (i.e., optical mode is not affected). Finally, after the glue is solidified, a vertical upward force is applied to detach the toroid from its silicon pillar. The toroid attached to the fiber tip can now be freely moved. This method is different from the one in Ref. \cite{hos2007} which uses a microfork. Note that Ref. \cite{hos2007} does not study coupled resonators.
\begin{figure}
\center \includegraphics[width=12cm]{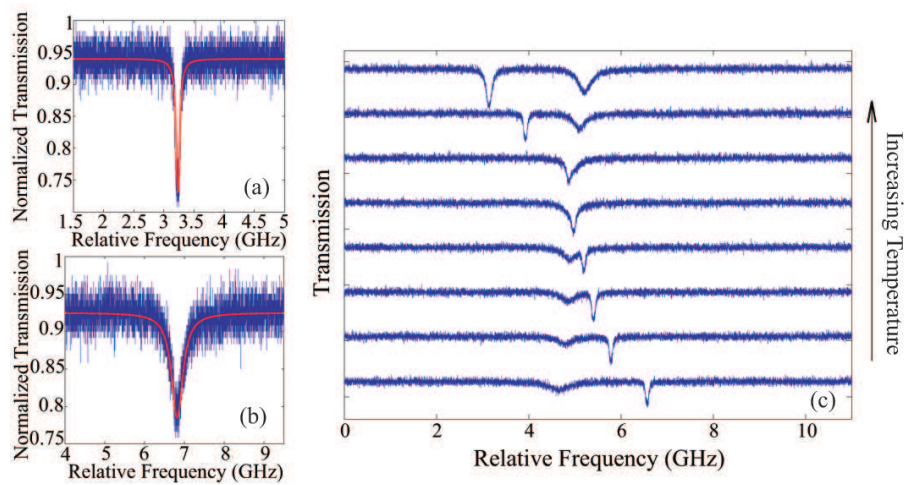} \caption{Thermal tuning of selected WGM modes of the resonators in PM2. Normalized transmission spectra obtained for (a) silica microsphere and (b) PDMS coated silica microtoroid. (c) Series of normalized transmission spectra taken in $1550 {\rm nm}$ band, showing thermal tuning.\vspace{-1 mm}} \label{Fig2}\end{figure}

For the excitation and characterization of the PMs, optical power from a tunable laser ($1550 {\rm nm}$ band) was coupled via fiber taper in and out of only one of the WGM resonators (on-chip microtoroid in PM1 and microsphere in PM2) forming the PM (Fig. \ref{Fig1}). The optical power was coupled to the on-chip microtoroid in PM1, and to the microsphere in PM2. The distance between the resonators of each PM and that between the fiber taper and the resonators were finely tuned using a 3D nano positioning system. The transmitted power through the fiber taper was monitored and analyzed as the wavelength of the laser was continuously scanned.

Each resonator in the PMs supports many optical modes; however, only the modes having spectral overlap with those in the other resonator contribute to the physical system of the PM. For the fiber coupled system, we consider the fundamental mode (the azimuthal mode located in the circular rim of the torus, with very high Q) \cite{yang2009}. The polarization state of the resonant mode is controlled by the polarization controller (Fig. \ref{Fig1}). The WGMs in the resonators of each PM is in general detuned from each other. However, they can be made frequency-degenerate by thermally tuning the resonance frequencies (e.g., by a thermo-electric cooler). As the temperature is decreased, the resonance frequencies of the silica resonators should experience blue shift while those in the PDMS coated silica microtoroid may experience blue-, red- or no-shift depending on the PDMS thickness \cite{he2009}. With the proper choice of thickness, we can make the PDMS coated microtoroid experience red-shift with decreasing temperature. This opposing direction of resonance shifts in silica and PDMS coated silica resonators are exploited to tune the modes and make them degenerate in frequency. Thus, we can drive the resonators of a PM from off-resonance (detuned or non-degenerate) to on-resonance (overlap or degenerate).
\begin{figure}
\center \includegraphics[width=14cm]{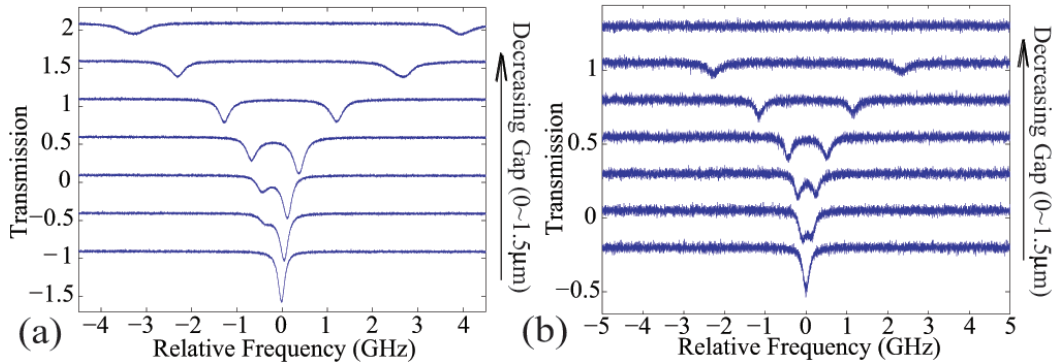} \caption{Transmission spectra of (a) PM1 and (b) PM2 as the distance between the resonators forming each PM is decreased from $1.5{\rm \mu m}$ with a step-resolution of $100 {\rm nm}$.\vspace{-1 mm}} \label{Fig3}\end{figure}

We chose one mode from each resonator such that their frequency detuning was within our thermal tuning range of $4 {\rm GHz}$. Initially, the resonators in PM1 had resonance wavelengths of $1554 {\rm nm}$ and $Q\sim 10^6$ (free toroid) and $Q\sim 2.1\times10^6$ (on-chip toroid), and those in PM2 had $1541.17{\rm nm}$ with $Q\sim10^6$ (PDMS coated on-chip toroid) and $1541.19{\rm nm}$  with $Q\sim3\times10^6$ (microsphere). These modes correspond to ${TE}^1_{95}$ and ${TE}^1_{82}$, respectively for PM1 and PM2. Since the WGMs of the two silica microtoroids in PM1 were frequency-degenerate, thermal tuning was not needed. In PM2, on the other hand,  initially the WGMs were non-degenerate and thermal tuning was employed. Figure \ref{Fig2}(a) and \ref{Fig2}(b) show the spectra of the modes of the resonators in PM2 before thermal tuning was applied. By thermal tuning, the chosen WGMs of the microsphere and the microtoroid in PM2 were tuned from off-resonance to on-resonance. The selected WGM of the PDMS coated toroid experienced a small red shift whereas that of the microsphere a blue shift, moving closer to the microtorid WGM as the temperature decreased (Fig. \ref{Fig2}(c)). At a certain temperature, the modes became degenerate. Further decrease shifted the microsphere mode further to blue side of the spectrum, crossing the WGM of the toroid and moving away.

\begin{figure}
\center \includegraphics[width=14cm]{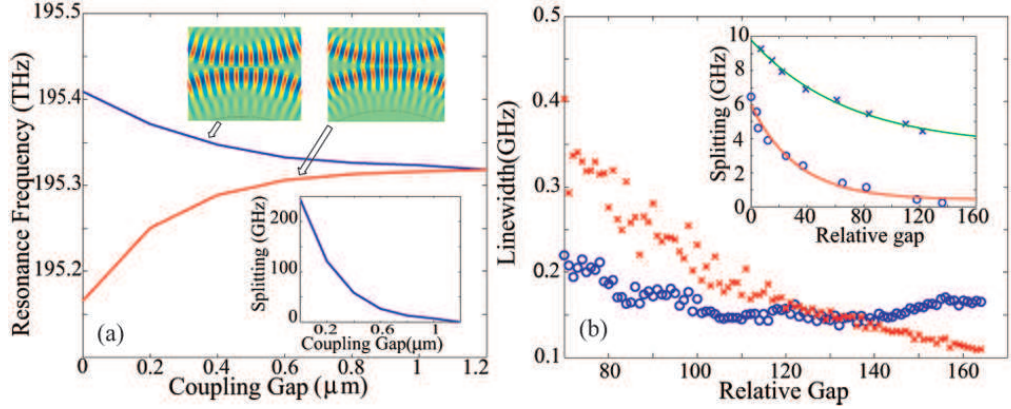} \caption{ (a) Simulation results showing the resonance frequencies of the supermodes and the splitting (inset) as a function of gap between the resonators. Also shown in the inset is the field distributions of symmetric and asymmetric supermodes. (b) Experimental result showing linewidth crossing. Inset is experimental confirmation of the exponential dependence of splitting on the gap between the resonators in PM1 (*) and PM2 (o). \vspace{-1 mm}} \label{Fig4}\end{figure}

After spectrally overlapping the WGM resonances of the two resonators of a PM, we tuned the gap between them to control the direct coupling and interaction of the degenerate modes. As soon as the direct coupling was achieved, two normal supermodes were formed (Fig. \ref{Fig3}). The field distributions of the split modes were $\pi$ shifted from each other, and were identified as symmetric and anti-symmetric modes as verified with numerical simulations carried out in COMSOL (Fig. \ref{Fig4}(a)). The spectral separation (splitting) of the modes depends on the distance between the resonators forming the PM. Figure \ref{Fig3} depicts the splitting spectra observed in the transmitted power through the fiber taper as the wavelength of the tunable laser is scanned while the gap between the resonators are continuously changed. Control of splitting frequency in the range $0-8 {\rm GHz}$ and $0-4 {\rm GHz}$ was demonstrated, respectively for the PM1 and PM2.

Splitting exponentially decreases with increasing distance between the resonators (Fig. \ref{Fig4}). Although numerical simulations suggest that splitting is maximized when the distance is zero, in the experiments we find that if the gap between the resonators is smaller than a certain value, the coupling of the supermodes become so strong that their mode profiles are significantly disturbed leading to $Q$-value reduction due to increased scattering losses. The resonances become shallower, and when the distance is close to zero, resonances are no longer observed (Fig. \ref{Fig3}(b)). In addition to splitting, linewidth crossing was also observed (Fig. \ref{Fig4}(b)).

Assuming a coupling strength $g$, which depends on the distance between the resonators and on how well their evanescent fields overlap, mode coupling is modeled as
\begin{eqnarray}
\frac{da_1}{dt} &=& \alpha_1a_1-iga_2-\sqrt{\kappa_c}a_{\rm in}\\
\frac{da_2}{dt} &=& \alpha_2a_2-iga_1
\end{eqnarray}
together with the input-output relation $a_{\rm out}=a_{\rm in}+\sqrt{\kappa_c}a_1$. In Eqs.1 and 2, $a_{\rm in}$ and $a_{\rm out}$  are the amplitudes of the input and output modes, respectively, $\kappa_c$ is the coupling rate of the mode between the taper and the resonator 1, $\omega_{1(2)}$, $a_{1(2)}$ and $\kappa_{1(2)}$ denote the resonance angular frequency, the mode amplitude and the dissipation rate of the mode in resonator 1(2), respectively. $\alpha_{1(2)}=i\Delta_{1(2)}-\kappa_{1(2)}/2$ with $\Delta_{1(2)}$ denotes the frequency detuning between the input light and the WGM resonances. We find the transmission at steady state as
\begin{eqnarray}
T=|1+\frac{\kappa_c\alpha_2}{\alpha_1\alpha_2+g^2}|^2
\end{eqnarray}
and the eigenfrequencies $\omega_{\pm}$ of the supermodes as
\begin{eqnarray}
\omega_{\pm}=\frac{i(\alpha'_1+\alpha'_2)}{2}\pm\sqrt{g^2-(\alpha'_1-\alpha'_2)^2/4}.
\end{eqnarray} where $\alpha'_{1(2)}=-i\omega_{1(2)}-\kappa_{1(2)}/2$. Then the splitting $\delta=|\omega_+-\omega_-|$ between the eigenfrequencies becomes $\delta=\sqrt{4g^2-(\alpha'_1-\alpha'_2)^2}$. When the resonances of the uncoupled resonators are tuned to be degenerate (i.e., $\omega_1=\omega_2$), we obtain $\delta_{(\omega_1=\omega_2)}=2^{-1}\sqrt{16g^2-(\kappa_1-\kappa_2)^2}$. For $g<|\kappa_1-\kappa_2|/4$, $\delta$ becomes imaginary and frequency splitting disappears, implying weak coupling where damping and hence the linewidths of the modes of the resonators are modified. For $g>|\kappa_1-\kappa_2|/4$, we have real $\delta>0$ implying splitting and strong coupling where the photons coherently transfers between the resonators forming the PM. As $g$ increases (i.e., gap between the resonators decreases), $\delta$ increases, too. This is consistent
with our experimental observations (Fig. \ref{Fig3}).

In conclusion, we have demonstrated coupling of on-chip and free resonators of different shapes and materials to form PMs. By thermally tuning the resonances of on-chip resonators, we achieved a good spectral overlap of the selected WGMs of the resonators in each PM. Clear mode splitting and linewidth crossings were observed in the transmission spectra as a function of the distance between the resonators of a photonic molecule.

\textbf{Acknowledgement} This work is supported by the National Science Foundation under grant No. 0954941.

\end{document}